\documentclass[prl,aps,twocolumn]{revtex4}
\usepackage{graphicx}
\begin{document}
\def \tr{{\mbox{tr~}}}
\def \ra{{\rightarrow}}
\def \ua{{\uparrow}}
\def \da{{\downarrow}}
\def \be{\begin{equation}}
\def \ee{\end{equation}}
\def \bea{\begin{eqnarray}}
\def \eea{\end{eqnarray}}
\def \nn{\nonumber}
\def \half{{1\over 2}}
\def \etal{{\it {et al}}}
\def \cH{{\cal{H}}}
\def \cM{{\cal{M}}}
\def \cN{{\cal{N}}}
\def \cQ{{\cal Q}}
\def \cI{{\cal I}}
\def \cV{{\cal V}}
\def \cG{{\cal G}}
\def \bS{{\bf S}}
\def \bL{{\bf L}}
\def \bG{{\bf G}}
\def \bQ{{\bf Q}}
\def \bR{{\bf R}}
\def \br{{\bf r}}
\def \bu{{\bf u}}
\def \bq{{\bf q}}
\def \bk{{\bf k}}
\def \tJ{{\tilde{J}}}
\def \W{{\Omega}}
\def \e{{\epsilon}}
\def \lam{{\lambda}}
\def \a{{\alpha}}
\def \t{{\theta}}
\def \b{{\beta}}
\def \g{{\gamma}}
\def \D{{\Delta}}
\def \d{{\delta}}
\def \w{{\omega}}
\def \s{{\sigma}}
\def \f{{\varphi}}
\def \x{{\chi}}
\def \h{{\eta}}
\def \hatt{{\hat{\t}}}
\def \hn{{\bar{n}}}
\def \vk{{\bf{k}}}
\def \vq{{\bf{q}}}
\def \gk{{\g_{\vk}}}
\def \nd{{^{\vphantom{\dagger}}}}
\def \yd{^\dagger}
\def \ket#1{{\,|\,#1\,\rangle\,}}
\def \bra#1{{\,\langle\,#1\,|\,}}
\def \braket#1#2{{\,\langle\,#1\,|\,#2\,\rangle\,}}
\def \expect#1#2#3{{\,\langle\,#1\,|\,#2\,|\,#3\,\rangle\,}}
\def \rl#1#2{{\,\langle\,#1\,#2\,\rangle\,}}
\def \ad{{\a\yd}}
\def \an{{\a\nd}}
\def \av#1{{\langle#1\rangle}}
\def \bd#1{{(\sin\t\ad_{0#1}+\cos\t\ad_{1#1})}}
\def \bn#1{{(\sin\t\an_{0#1}+\cos\t\an_{1#1})}}
\def \sd#1{{(\cos\t\ad_{0#1}-\sin\t\ad_{1#1})}}
\def \sn#1{{(\cos\t\an_{0#1}-\sin\t\an_{1#1})}}

\title{Probing many-body states of
ultra-cold atoms via noise correlations }
\author{Ehud Altman, Eugene Demler, and Mikhail D.Lukin}
\address{Physics Department, Harvard University, Cambridge, MA 02138}
\date{June 9, 2003}
\begin{abstract}
We propose to utilize density-density correlations in the image of an expanding gas cloud
to probe complex many body states of trapped ultra-cold atoms.
In particular we show how this technique can be used to detect superfluidity
of fermionic gases and reveal broken spin symmetries in Mott-states of atoms
in optical lattices.
The feasibility of the method is investigated by
analysis of the relevant signal to noise ratio including experimental
imperfections.
\end{abstract}
\maketitle


Much of the excitement in the field of Bose-Einstein condensation
owes to the clear demonstration it provides of the wave character
of matter.
The condensed state of bosons involves
macroscopic occupation of a delocalized single particle state.
Consequently, it is characterized by sharp density peaks in the
freely expanding gas cloud after it is released from the trap\cite{BECexp}.
Patterns that appear when two or more superfluid clouds interfere\cite{andrews},
are a direct probe of the single particle coherence, amplified by macroscopic occupation.

Recent experiments open intriguing directions for studying many body phenomena
beyond single particle coherence. For example,
observation of the superfluid to Mott insulator transition
\cite{Greiner}, as well as experiments involving ultra-cold
fermions near a Feshbach resonance\cite{truscott},
address strongly correlated states of matter.
The most intriguing aspect of such systems is the existence of non-trivial correlations
and complex order that defy a description in terms of (single particle) matter waves.
Accordingly, they cannot be characterized simply by the density profile
of an expanding cloud. For example, the localized atoms in Mott states of the
optical lattice display a vanishing interference pattern
\cite{Greiner}, which can hardly reveal detailed properties of
the quantum state. Likewise, superfluidity of paired fermions is
not evident as a coherence peak in the density profile\cite{stoof},
and detecting the order parameter, presents a considerable challenge.
Observation of some theoretically proposed ``exotic'' many-atom states
\cite{zhou,duan,recati} may prove even more elusive.

In this Letter, we show that the quantum nature
of strongly correlated states can be revealed
by {\em spatial noise} correlations in the image of the expanding gas.
This is similar in spirit to measurements of non classical correlations
of light in optical systems \cite{book} and {\em temporal}
current noise in mesoscopic conductors\cite{noise-review}.
In analogy to quantum optics, this technique allows to study
matter waves that lack single particle coherence.
Specifically we show that
(i) fermionic atoms released from the trap would display a clear
signature of superfluidty in their density correlations.
Furthermore, detailed properties of the fermionic
superfluidity can be determined,
such as pairing symmetry, and BCS to BEC crossover\cite{randeria};
(ii) Atoms released
from a Mott-insulating state of the optical lattice display sharp
(Bragg) peaks in the
density-density correlation function as a
consequence of quantum statistics;
(iii) These peaks
can be used to probe the spin ordered Mott states proposed for
two component bosons \cite{kuklov,duan}.
Finally we verify the experimental feasibility of the proposed measurements.

Before proceeding we note that earlier proposals
to detect fermionic superfluidity
relied on dynamical response of the cloud  \cite{stringari},
inelastic scattering of light to induce and measure
excitations\cite{walter,zoller-old}, or to
microscopically image the pair wave-function in the trap\cite{zhang}. In contrast,
the present technique provides a direct probe
of the pair coherence as well as information on the pairing symmetry. Earlier work on
probing Mott-insulating phases should also be noted \cite{roberts}.
\begin{figure}[t]
  \centering
  \includegraphics[width=8cm]{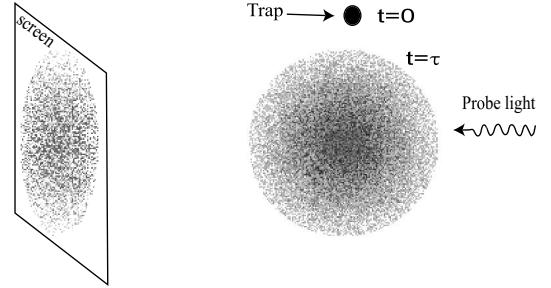}
\caption{Time of flight imaging. The atoms are released
from the confining potential at time $t=0$ and the density of the expanding
cloud is imaged at a later time. Spatial noise correlations
in this image can be used to probe the quantum state at $t=0$.}
\label{fig:setup}
\end{figure}

We proceed by formulating a detection scheme for atoms
released from a single macroscopic trap.
Suppose, for simplicity, that the system is initially in some pure state
$\ket{\Phi}$. In a typical experimental setup (Fig. \ref{fig:setup}),
the trapping potential is
turned off suddenly, and  the atoms evolve {\it independently} under
the influence of the free propagator $U_0(t)$. This is valid provided that
the free-atom collision cross-section is not too large. Such conditions can be achieved
by switching the magnetic field to values far from the Feshbach resonance when turning off the trap\cite{truscott}.

In such time of flight experiments, the column integrated density
of the expanding cloud is measured by light absorption imaging\cite{Ketterle-Varena}.
The images are commonly
analyzed by comparing to theoretical predictions for the
density expectation value:
\be
\av{\hat{n}_\a(\br)}_t=\bra{\Phi}U_0\yd(t)\psi_\alpha\yd(\br)\psi_\alpha\nd(\br)U_0(t)
\ket{\Phi}
\label{g1}
\ee
where $\psi_\alpha$ is the field operator for bosons or fermions and
$\alpha$ denotes an internal atomic quantum number (spin).
After a long time of flight the density distribution becomes proportional
to the momentum distribution in the initial trapped state
$\av{n(\br)}_t\approx(m/ht)\av{{\hat n}_{\bQ(\br)}}$. The wave-vector
$\bQ(\br)=m\br/(\hbar t)$
defines a correspondence between
position in the cloud and momentum in the trap.

It is important to realize, that in each experimental image, a
single realization of the density is observed, not the expectation
value. Eq. (\ref{g1}) is still meaningful, because the density is
a self averaging quantity. Each bin $\s$ in the image represents a
substantial number of atoms $N_\s$, while the atomic noise scales
as $O(\sqrt{N_\s})$. But since $N_\s$ is not macroscopic the
density fluctuations are visible. They are characterized by the
correlation function: \be \cG_{\alpha\beta}(\br,\br')=
\av{\hat{n}_\alpha(\br)\hat{n}_\beta(\br')}_t -
\av{\hat{n}_\alpha(\br)}_t\av{\hat{n}_\beta(\br')}_t. \ee In
analogy with Eq. (\ref{g1}) this can be related to ground state
momentum correlations: \bea \cG_{\alpha\beta}(\br,\br')\propto
\av{\hat{n}_{\bQ(\br)\a} \hat{n}_{\bQ(\br')\b}}
-\av{\hat{n}_{\bQ(\br)\a}}\av{\hat{n}_{\bQ(\br')\b}}, \label{dd}
\eea The proportionality constant is $(m/\hbar t)^6$ and we shall
omit it henceforth. We shall be concerned primarily with pure
density-density correlations
$\cG(\br,\br')=\sum_{\a\b}\cG_{\alpha\beta}(\br,\br')$, which does
not require state-selective measurement. In practice it may be
more convenient to consider the quantity $\D n(\br,\br')\equiv
n(\br)-n(\br')$ whose fluctuations are closely related to
$\cG(\br,\br')$. If $\av{n(\br)}_t=\av{n(\br')}_t$, then \bea
\av{\D n(\br,\br')^2}_t= \cG(\br,\br) + \cG(\br',\br') -
2\cG(\br,\br'). \label{g2} \eea

{\em Fermionic superfluids.} As a specific example, we now consider
superfluid states of fermionic atoms.
Such superfluids sustain
macroscopic coherence and their transport properties are similar to their
bosonic counterparts. However, the first order coherence measured by the
density profile of the expanding cloud would not reveal their superfluid nature.
Consider a system at zero temperature described by
a BCS-like ground state
\be
\ket{\Phi_{BCS}}=\prod_k (u_{\bk}+v_{\bk}a\yd_{\bk\ua}a\yd_{-\bk\da})\ket{0}.
\label{BCS}
\ee
Note that this wave function can describe
both weak-coupling BCS-like pairing, as well as tightly bound pairs for which
$u_{\bk}$ and $v_{\bk}$ have a wide momentum distribution\cite{randeria}.
The average density profile of the expanding cloud (\ref{g1}) is
proportional to the BCS momentum distribution
function
$\av{\hat{n}(\br)}_t=2|v_{\bQ(\br)}|^2$,
which is qualitatively indistinguishable
from a Fermi distribution at $T=T_C$ \cite{truscott}.

The essential difference between these states lies in the two particle correlations.
For every atom with momentum $\bk$ in the BCS state,
there is another one at exactly $-\bk$. This implies,
pronounced correlations
between density fluctuations on diametrically opposite points in the expanding cloud.
Indeed, a straightforward application of (\ref{dd}) for the BCS state gives:
\be
\cG(\br,\br')= 2|u_{\bQ(\br)}|^2|v_{\bQ(\br')}|^2\tilde{\d}(\br+\br'),
\label{Gbcs}
\ee
where $\tilde{\d}(\br+\br')$ is a sharply peaked function of $\br+\br'$. This sharp
peak is a direct analogue of the
{\em first order} coherence peak seen in bosonic condensates. It
indicates condensation of zero momentum pairs. As for BEC, it has a width
$\sim\hbar t/mL$, limited by the finite size $L$ of the initial cloud. The pair distribution
$n_s(\br)=2|u_{\bQ(\br)}|^2|v_{\bQ(\br)}|^2$ can be read off directly from the weight of the peak.
\begin{figure}[h]
  \centering
  \includegraphics[height=6cm,width=8cm]{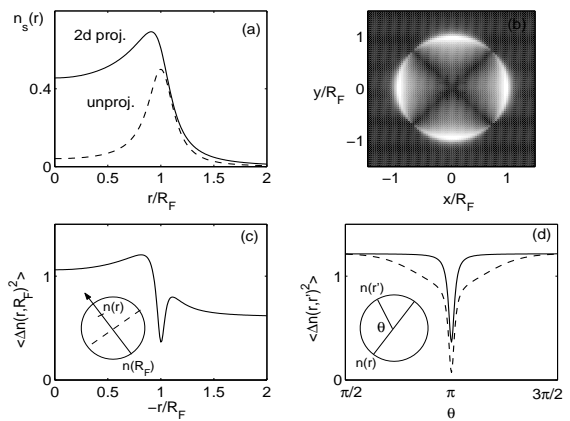}
\caption{{\em Fermionic superfluid.} (a)
Angle integrated weight of the correlation peak at diametrically
opposite points. (b) Weight of the correlation peak for $d_{x^2-y^2}$ pairing. (c) and (d)
$\av{\D n(\br,\br')^2}$ at $T>0$ (superfluid fraction 0.7).
The width of the narrow dip $\sim \hbar t/(mL)$ is limited by the system's size.
The dashed line in (d): BEC of tightly bound pairs. }
\label{fig:bcs}
\end{figure}

Fig \ref{fig:bcs}(a) depicts the angular averaged $n_s(r)$ (dashed). The solid line as well
and other plots in Fig \ref{fig:bcs} depict observable column integrated functions.
Pairing symmetries other than $s$-wave, would be detected by higher angular
harmonics in the pair distribution $n_s(\br)$ and its column integral.
As an example we plot this function for $d_{x^2-y^2}$ pairing in Fig. \ref{fig:bcs}(b).

In practice it may be more convenient to measure the fluctuations
of $\D n(\br,\br')$. Noting that the state (\ref{BCS}) is an eigenstate
of $\D n(\br,-\br)$ with eigenvalue zero, we have $\av{\D n(\br,-\br)^2}\equiv 0$ at $T=0$.
Well
away from $\br+\br' =0$
$\langle \D n(\br,\br')^2\rangle \sim n_{\bQ(\br)}(1-n_{\bQ(\br)})+n_{\bQ(\br')}(1-n_{\bQ(\br')})$,
which is non-zero on the smeared fermi surface of a fermi liquid (at $T>0$) and for a BCS state.
A sharp dip, whose width $\hbar t/(mL)$ is
limited by the system size, appears around $\br'=\br$ in the superfluid state (Fig \ref{fig:bcs}(c,d)).
Note that at finite $T$ the magnitude of the dip is reduced and it disappears at $T_C$.
Heating a tightly bound pair condensate
will produce thermal pairs at momenta $k>0$, producing a wide background dip
in analogy to the bimodal distribution in BEC.

{\em Optical lattice.} We now consider atoms initially confined to an optical lattice.
Calculation of $\av{n(\br)}$ and $\av{n(\br)n(\br')}$
involves operating on the lattice state $\ket{\Phi}$ with
$U(t)\yd\psi(\br)U(t)$. In normal ordered expectation values this operator can be
safely replaced by its projection into the lowest Bloch band,
$A(\br,t)\equiv\sum_i w_i(\br,t)a_{i\a}$ \cite{roth}.
Here  $w_i(\br,t)$ is the free
evolution of a Wannier wave function, initially
centered around the lattice site at $R_i$ and
$a_{i\alpha}=\int d\br w_i(\br,0)\psi_{\alpha}(\br)$ is a particle annihilation operator
at this site. At long times $w_i(\br,t)\propto e^{-i\bQ(\br)\cdot\bR_i}$
times a gaussian envelope,which we shall approximate by a square of the same width $W=\hbar t/(a_0 m)$.
$a_0$ is the
width of the Wannier state on the lattice.
Now $\av{\hat{n_{\alpha}}(\br)}_t$ can be related
to the off-diagonal
correlation function in the ground state\cite{roth}:
\be
\av{\hat{n_\a}(\br)}_t \approx \av{\hat{n}_{\bQ(\br)}} \equiv
\frac{1}{W^d}\sum_{i,j} e^{i(\bR_i-\bR_j)\cdot\bQ(\br)}
\av{a\yd_{i\a} a\nd_{j\a}},
\label{g1p}
\ee
The wave-vector $\bQ(\br)$ now defines a correspondence between
position in the cloud and the quasi-momentum on the lattice.
$\cG$ can also be
written in terms of correlations in the lattice state:
\bea
\cG_{\a,\b}(\br,\br')&\sim & \sum_{ii'jj'}e^{i\bR_{ii'}\cdot\bQ(\br)
+i\bR_{jj'}\cdot\bQ(\br')}
\av{a\yd_{i\a} a\yd_{j \b} a\nd_{j'\b} a\nd_{i'\a}}\nn\\
&&+\d(\br-\br')\av{n(\br)}_t-\av{n(\br)}\av{n(\br')}
\label{ddl}
\eea
Here $\d(\br-\br')$ is a true $\d$-function originating from
normal ordering in the continuum.
Note that formulas (\ref{g1p}) and (\ref{ddl}) hold for bosons and fermions.

In the superfluid state of bosons, where $\av{a\yd_i a\nd_j}=|\Psi|^2$,
$n(\br,t)$ exhibits
Bragg peaks at $\bQ(\br)$ corresponding to reciprocal lattice vectors
$\bG$.
In the Mott state on the other hand $\av{a\yd_i a\nd_j}\approx\d_{ij}$,
and there is no interference pattern in $n(\br,t)$.

The Mott state, however, displays non-trivial correlations in the second-order
correlation function (\ref{ddl}) associated with
atom number fluctuations. To illustrate their origin, and to make contact
with the Hanbury-Brown-Twiss effect in quantum optics, we first
apply (\ref{g1p},\ref{ddl}) to the "toy" model of two atoms localized in two
 wells.
For an initial Fock state $\ket{\Phi}=a\yd_1a\yd_2\ket{0}$ it gives:
\be
\cG(\br,\br')= \frac{2}{W^{2d}}\h\cos\left(\frac{m}{\hbar t}(\br-\br')\right)
\label{bunching}
\ee
where $\h=1~(-1)$ for bosons
(fermions). The oscillations reflect a two body interference effect which
amounts to bunching in the case of bosons and anti-bunching for fermions, as
in the textbook example\cite{book,yamamoto}.

Addressing the optical lattice we first consider
the two extremes
of a superfluid and a Mott state:
\bea
\ket{SF}=\prod_i e^{z a\yd_i}\ket{0}, \ket{Mott}=\prod_i
\frac{1}{\sqrt{n!}}(a\yd_i)^n\ket{0}.
\label{B-WF}
\eea
According to Eq. (\ref{ddl}), $\cG_{SF}(\br,\br')\equiv 0$, while
\bea
\cG_{Mott}(\br,\br')\approx \frac{N}{W^d}\left(\frac{2\pi a_0}{l}\right)^d\sum_\bG
\tilde{\d}^d\left(\br-\br'+\frac{\hbar t}{m}\bG\right)
\eea
where $l$ is a lattice spacing and $a_0$, the width of a Wannier function.
We conclude that the Mott state sustains second order coherence,
seen as Bragg peaks in $\cG(\br,\br')$.
Note, again,  that for a finite lattice of length $L$, the peaks are not true $\d$-functions.
They have a finite hight $(N/W^d)^2$ and width $\xi=2\pi\hbar t/mL$.

\begin{figure}[h]
  \centering
  \includegraphics[width=6cm]{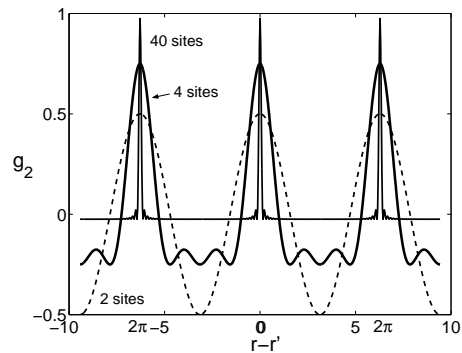}
\caption{Normalized density density correlations,
$g\equiv \cG(r,r')/\av{n(r)}_t\av{n(r')}_t$ in the Mott state for
a chain of 2,4 and 40 sites.} \label{fig:mott}
\end{figure}

Fig. \ref{fig:mott} shows how the two-site bunching signature
of Eq. (\ref{bunching})
develops into coherent Bragg peaks with increasing lattice size.
A fermionic insulating state will display similar evolution, only the
peaks will be replaced by dips.

Note that the Bragg peaks do not reflect reduced
number fluctuations in the Mott state, but rather the quantum statistics
of bosonic or fermionic atoms. In particular, we note that
a thermal Bose gas on a lattice would
exhibit similar coherent Bragg peaks, which can be seen
as a result of the permutation
symmetry imposed on the wave-function. However in a Mott state
with non trivial internal structure, the noise correlations
can reveal complex order as we now illustrate.

As a specific example let us consider
spin ordered Mott states, that have been predicted in the Mott phase
of two component bosons on an optical lattice \cite{duan}.
The two components can be represented by a
pseudospin-$\half$: $a_{\ua (\da)}\yd\ket{0}\equiv\ket
{\ua (\da)}$.
A fairly general mean-field state of
a spin-ordered
insulator is given by the wave-function
\be
\ket{\Psi_{\bf q}}=\prod_i
(\cos\frac{\t}{2}a_{i\ua}\yd+e^{i\bq\cdot \bR_i}\sin\frac{\t}{2}a_{i\da}\yd)\ket{0}.
\label{Psi_q}
\ee
This state has a non zero spin component along the ${\hat z}$ axis and
helical component with momentum $q$ on the $xy$ plane. Note that the
limiting case $q=0$ corresponds to a $xy$ ferromagnet  while $q =\pi$
is a $Ne\acute{e}l$ state.
An application of (\ref{ddl}) to the state (\ref{Psi_q}), yields after summation
over spin indices:
\bea
&&\cG(\br,\br')\propto
\left(\cos^4\frac{\t}{2}+\sin^4\frac{\t}{2}\right)
\sum_\bG\tilde{\d}^d\Big(\br-\br'+\frac{\hbar t}{m}\bG\Big)\nn\\
&&+\sin^2\t\sum_\bG\sum_{\sigma=\pm}\tilde{\d}^d
\Big(\br-\br'+\frac{\hbar t}{m}(\sigma\bq+\bG)\Big).
\label{sidepeaks}
\eea

We find that the $\bq\ne 0$ spin order produces satellite peaks, whose separation
from the central peaks at reciprocal lattice vectors is proportional to $\pm q$. The
intensity ratio of the satellite to the central peaks depends on
the $\hat{z}$ polarization of the spins. The satellites
reach their maximal relative intensity, half that of the central peaks,
for spins polarized on the $xy$ plane. Note that in this scheme, an $x-y$
ferromagnet cannot be observed directly. However a time dependent helical twist
can be induced to this state by a magnetic field gradient.

{\em Experimental Issues.} To observe the proposed effects,
two criteria must be met:
(i) the atomic noise should be
observable in an experiment and (ii) correlated fluctuations should rise above the uncorrelated
noise in a statistical sense.

In a typical experiment, (\ref{Gbcs}) would be averaged over narrow cylinders
around $\br$ and $\br'$, whose bases $\s$ correspond to the spatial resolution.
The noise of detected probe-photons in a bin of area $\s$
is a sum of contributions from
atomic and laser light fluctuations. The later is fundamentally limited by
photon shot noise. The atomic
noise exceeds the photon shot noise, provided the number of photons per bin in the incoming pulse,
$p_{\s} > \exp(2\kappa)\langle  N_\s\rangle/(\eta \kappa^2)$, where $\kappa$ is the optical depth of the cloud,
which should be chosen close to unity,  and $\eta$ is the {\em photon} detection efficiency.
The practical limitation on the number of photons
is associated with photon recoil which results
in image blurring. Under standard conditions $10-100$ photons per atom can be
scattered in a measurement time of $10\mu s$
without blurring the image on the scale of $10 \mu m$ \cite{Ketterle-Varena},
indicating that detection of atomic noise is possible.
A typical cloud of $N = 10^7$ $^6Li$ atoms \cite{ket}
 will reach optical depth $\kappa\sim 1$
after expansion to about $1 m m$. Therefore each
$10 \mu m$  bin will
contain about $10^3$ atoms. The atomic noise contribution will cause bin-to-bin
variation in the optical absorption at the level
of few percent, which is detectable with current technology.

We now verify that in the case of a fermionic superfluid the
correlated fluctuations rise above the background.
For fixed $\br$ the value of $\D n(\br,\br')^2$ will exhibit fluctuations of order $n_s(r)$
as $\br'$ varied. To observe the superfluid dip at $\br+\br'=0$  we must average the result
over a number of independent positions $\br$. Different positions yield independent results if
they correspond to distinct momenta in the trap. Momentum quantization thus limits
the number of independent positions on the 2d image of the cloud to $O(N^{2/3})$,
where $N$ is the total number of atoms. Consequently the background statistical fluctuations
can be averaged to $~N^{-1/3}$ of the dip magnitude making it
observable in a single image measurement.

In conclusion, we demonstrated that spatial noise correlations in the image
of an expanding gas cloud released from a trap, can reveal key properties
of strongly correlated states of cold atoms. We anticipate that a similar
technique can be used to observe signatures of more exotic
states such as spin liquids and valence bond solids.

We acknowledge   valuable discussions with I. Bloch, M.Prentiss and W. Ketterle . This work was supported by
ARO, NSF (PHY-0134776, DMR-0132874), Sloan Foundation and Packard Foundation.

\end{document}